# Quantum Entanglement and the Generalized Uncertainty Principle

**Gardo Blado[*], Francisco Herrera, and Joshuah Erwin**

*Department of Mathematics and Physics, College of Science and Mathematics*

*Houston Baptist University*

*7502 Fondren Rd., Houston, Texas, U.S.A*

## Abstract

We examine quantum gravity effects on entanglement by a straightforward application of the generalized uncertainty principle (GUP) to continuous-variable systems. In particular, we study the following cases: the modified uncertainty relation of two identical entangled particles (Rigolin, 2002), and the inseparability conditions for entangled particles in the bipartite (Duan, Giedke, Cirac and Zoller, 2000) and tripartite (van Loock and Furusawa, 2003) cases. Rigolin showed a decrease in the lower bound of the product of the uncertainties of the position and momentum for two identical entangled particles while Duan and van Loock derived inseparability conditions for EPR-like operators. In all three cases, the GUP correction resulted in a higher value of the bounds: a higher lower bound for the Rigolin's result and a higher upper bound for the inseparability condition in Duan and van Loock's relations. In Rigolin's case, the GUP correction decreased the disagreement with the Heisenberg uncertainty relation while in Duan's and Loock's case, the inseparability and entanglement conditions are enhanced. Interestingly, the GUP corrections tend to make quantum mechanical effects more pronounced.

[*] Corresponding author:
*Email address*: gblado@hbu.edu

# 1. Introduction

The formulation of a viable quantum gravity theory has long been sought by theoretical physicists as part of the effort to unify the four forces of nature: electromagnetic, weak, strong and gravitational forces. An interesting phenomenological consequence of quantum gravity theories (like string theory and loop quantum gravity) is the presence of a minimal length scale which can result from the modification of the Heisenberg uncertainty principle (HUP) to a generalized uncertainty principle (GUP) [1-11]. The HUP, $\Delta x \Delta p \geq \hbar/2$ is changed to [9]

$$\Delta x \Delta p \geq \frac{\hbar}{2}[1 + \beta(\Delta p)^2 + \beta\langle p\rangle^2]$$

due to the modification of the commutator $[x,p] = i\hbar$ to

*Equation 1:* $[x,p] = i\hbar[1 + \beta p^2]$.

It can be shown that keeping the position operator the same while changing the momentum operator as

*Equation 2:* $x = x_0, p = p_0(1 + \beta p_0^2)$

in which $x_0$ and $p_0$ satisfy the HUP commutator, $[x_0, p_0] = i\hbar$, can give rise to the GUP. $\beta$ is called the GUP parameter which is small. Equation 2 leads to a modified Schrodinger equation $(p_0^2/2m + \beta p_0^4/m + V)\psi = E\psi$ to order $\beta$ with $p_0 = (\hbar/i)\, d/dx$. This has led to a number of phenomenological studies on GUP effects [1, 3, 5-9, 12-13].

In this paper we will study the GUP effects on quantum entanglement by applying the GUP commutator in Equation 1 to the entanglement-modified uncertainty relations derived by Rigolin [14-15], and the inseparability conditions calculated for the bipartite system by Duan [16] and the tripartite system by Loock [17]. In section 2, we discuss the connection of entanglement and the uncertainty relations through the work of Rigolin, Duan and Loock. The essential derivations are reproduced to make the calculation of the GUP corrections in section 3 apparent. We give our conclusions in section 4.

## 2. Entanglement and the Heisenberg Uncertainty Principle

Entanglement and the uncertainty principle can be related through the HUP or variance relations [14-22] and the entropic uncertainty principle [23-25]. In this paper, we will only study the HUP expressed as a variance relation.

### 2.1. Modified Uncertainty Relations for Identical, Entangled Particles

Rigolin's papers [14, 15] derived a new uncertainty relation involving an ensemble of $N$ identical and entangled particles. Using only physical observables and non-factorizable states for $N$ identical and entangled particles one can define the position and momentum physical observables in one dimension as,

*Equation 3:* $Q = Q_1 + Q_2 + \cdots + Q_N$ and $P = P_1 + P_2 + \cdots + P_N$

where $Q_i$ and $P_i$ are the position and momentum observables of the *i*th particle. From the uncertainty relation [27] $(\Delta A)^2 (\Delta B)^2 \geq \left(\frac{1}{2i}\langle [A,B] \rangle\right)^2$, we get

*Equation 4:* $(\Delta Q)^2 (\Delta P)^2 \geq \left(\frac{1}{2i}\langle [Q,P] \rangle\right)^2$.

With the usual commutation relations,

*Equation 5:* $[Q_i, P_j] = i\hbar \delta_{ij}$, $[Q_i, Q_j] = [P_i, P_j] = 0$,

it follows that

*Equation 6:* $[Q, P] = [Q_1, P_1] + [Q_2, P_2] + \cdots + [Q_N, P_N] = N(i\hbar)$.

Equation 4 becomes

*Equation 7:* $(\Delta Q)^2 (\Delta P)^2 \geq \frac{N^2 \hbar^2}{4}$

for N identical and entangled particles.

Let us consider two particles ($N = 2$). From

*Equation 8:* $(\Delta A)^2 = \langle A^2 \rangle - \langle A \rangle^2$,

we get

*Equation 9:* $(\Delta Q)^2 = \langle Q^2 \rangle - \langle Q \rangle^2$

where $Q$ for two particles in Equation 3 is $Q = Q_1 + Q_2$. Upon expansion of the right hand side of Equation 9, and with Equation 5, we get

*Equation 10:* $(\Delta Q)^2 = (\Delta Q_1)^2 + (\Delta Q_2)^2 + 2\langle Q_1 Q_2 \rangle - 2\langle Q_1 \rangle \langle Q_2 \rangle$

where we also used the definition in Equation 8. A similar calculation for the momentum operator yields

*Equation 11:* $(\Delta P)^2 = (\Delta P_1)^2 + (\Delta P_2)^2 + 2\langle P_1 P_2 \rangle - 2\langle P_1 \rangle \langle P_2 \rangle$.

Using Equation 10 and Equation 11 in Equation 7, for $N = 2$ particles, it follows that

*Equation 12:* $\left[\frac{1}{2}(\Delta Q_1)^2 + \frac{1}{2}(\Delta Q_2)^2 + \langle Q_1 Q_2 \rangle - \langle Q_1 \rangle \langle Q_2 \rangle\right] \times \left[\frac{1}{2}(\Delta P_1)^2 + \frac{1}{2}(\Delta P_2)^2 + \langle P_1 P_2 \rangle - \langle P_1 \rangle \langle P_2 \rangle\right] \geq \frac{\hbar^2}{4}$.

We need to rewrite the preceding equation entirely in terms of $\Delta Q_i$ and $\Delta P_i$. Let us consider the states $|\varphi_1\rangle = (Q_1 \pm Q_2)|\varphi_2\rangle$ and $|\varphi_2\rangle = |\psi\rangle$ in which $|\psi\rangle$ is a normalized two-particle state. We can use the Schwarz inequality in the form of $\langle \varphi_1 | \varphi_1 \rangle \langle \varphi_2 | \varphi_2 \rangle \geq \langle \varphi_1 | \varphi_2 \rangle \langle \varphi_2 | \varphi_1 \rangle$. Taking the $|\varphi_1\rangle =$

$(Q_1 + Q_2)|\varphi_2\rangle$ case, we get $\frac{1}{2}(\Delta Q_1)^2 + \frac{1}{2}(\Delta Q_2)^2 \geq -(\langle Q_1 Q_2 \rangle - \langle Q_1\rangle\langle Q_2\rangle)$ while taking the $|\varphi_1\rangle = (Q_1 - Q_2)|\varphi_2\rangle$ case, we get $\frac{1}{2}(\Delta Q_1)^2 + \frac{1}{2}(\Delta Q_2)^2 \geq +(\langle Q_1 Q_2 \rangle - \langle Q_1\rangle\langle Q_2\rangle)$, hence we get $\frac{1}{2}(\Delta Q_1)^2 + \frac{1}{2}(\Delta Q_2)^2 \geq |\langle Q_1 Q_2 \rangle - \langle Q_1\rangle\langle Q_2\rangle|$. Adding $\frac{1}{2}(\Delta Q_1)^2 + \frac{1}{2}(\Delta Q_2)^2$ on both sides of the preceding equation, we get $(\Delta Q_1)^2 + (\Delta Q_2)^2 \geq \frac{1}{2}(\Delta Q_1)^2 + \frac{1}{2}(\Delta Q_2)^2 + |\langle Q_1 Q_2 \rangle - \langle Q_1\rangle\langle Q_2\rangle| \geq \frac{1}{2}(\Delta Q_1)^2 + \frac{1}{2}(\Delta Q_2)^2 + \langle Q_1 Q_2 \rangle - \langle Q_1\rangle\langle Q_2\rangle$ in which the last inequality is true because of the absolute value sign. Summarizing, we have

*Equation 13:* $(\Delta Q_1)^2 + (\Delta Q_2)^2 \geq \frac{1}{2}(\Delta Q_1)^2 + \frac{1}{2}(\Delta Q_2)^2 + \langle Q_1 Q_2 \rangle - \langle Q_1\rangle\langle Q_2\rangle$.

Similarly, we get

*Equation 14:* $(\Delta P_1)^2 + (\Delta P_2)^2 \geq \frac{1}{2}(\Delta P_1)^2 + \frac{1}{2}(\Delta P_2)^2 + \langle P_1 P_2 \rangle - \langle P_1\rangle\langle P_2\rangle$.

From Equation 13 and Equation 14, we can rewrite the inequality Equation 12 entirely in terms of $\Delta Q_i$ and $\Delta P_i$ as

*Equation 15:* $[(\Delta Q_1)^2 + (\Delta Q_2)^2] \times [(\Delta P_1)^2 + (\Delta P_2)^2] \geq \frac{\hbar^2}{4}$.

This equation describes two identical, entangled particles. Considering the special case when $\Delta Q_1 = \Delta Q_2$ and $\Delta P_1 = \Delta P_2$, we get $(\Delta Q_i)^2(\Delta P_i)^2 \geq \frac{\hbar}{16}$ or

*Equation 16:* $\Delta Q_i \Delta P_i \geq \frac{\hbar}{4}, i = 1,2$.

Equation 16 implies a modified uncertainty relation for entangled particles. Equation 16 shows that entangled particles decrease the uncertainty of the position and momentum. This is consistent with reference [24] which showed that increased entanglement leads to less uncertainty.

### 2.2. Separability Condition for Entangled Particles, Bipartite Case

The paper by Duan et al. [16] derived a sufficient condition for inseparability for the entanglement of bipartite systems with continuous variable states using the HUP.

Consider the density operator $\rho$ (which is a composite bipartite state with modes 1 and 2) associated with a <u>separable</u> quantum state. This operator can be written as

*Equation 17:* $\rho = \sum_i \eta_i \rho_{i1} \otimes \rho_{i2}, \eta_i \geq 0$ and $\sum_i \eta_i = 1$

with $i = 1, 2, ...$ which labels the normalized states of the two modes. Duan et. al. constructed the EPR-like operators

*Equation 18:* $\hat{u} = |a|\hat{x}_1 + \frac{1}{a}\hat{x}_2$ and $\hat{v} = |a|\hat{p}_1 + \frac{1}{a}\hat{p}_2$

where $a$ is an arbitrary nonzero, real number and $\hat{x}_i$ and $\hat{p}_j$ obey the commutation relations in Equation 5, $[\hat{x}_i, \hat{p}_j] = i\delta_{ij}$, $i, j = 1,2$ where we set $\hbar = 1$. It follows from the uncertainty relation that

*Equation 19:* $\langle(\Delta\hat{x}_j)^2\rangle_i + \langle(\Delta\hat{p}_j)^2\rangle_i \geq |\langle[\hat{x}_j, \hat{p}_j]\rangle_i| = 1$ with $j = 1,2$

where the symbol $\langle\cdots\rangle_i$ denotes averaging over the product density operator $\rho_{i1} \otimes \rho_{i2}$. From Equation 8 and Equation 18, we get

*Equation 20:* $\langle(\Delta\hat{u})^2\rangle_\rho + \langle(\Delta\hat{v})^2\rangle_\rho = \Delta^2 + \sum_i \eta_i \langle\hat{u}\rangle_i^2 - (\sum_i \eta_i \langle\hat{u}\rangle_i)^2 + \sum_i \eta_i \langle\hat{v}\rangle_i^2 - (\sum_i \eta_i \langle\hat{v}\rangle_i)^2$

where we define for convenience

*Equation 21:* $\Delta^2 \equiv \sum_i \eta_i \left(a^2 \langle(\Delta\hat{x}_1)^2\rangle_i + \frac{1}{a^2}\langle(\Delta\hat{x}_2)^2\rangle_i + a^2\langle(\Delta\hat{p}_1)^2\rangle_i + \frac{1}{a^2}\langle(\Delta\hat{p}_2)^2\rangle_i\right)$.

Rewriting Equation 21, we get

$\Delta^2 \equiv \sum_i \eta_i \left(a^2 [\langle(\Delta\hat{x}_1)^2\rangle_i + \langle(\Delta\hat{p}_1)^2\rangle_i] + \frac{1}{a^2}[\langle(\Delta\hat{x}_2)^2\rangle_i + \langle(\Delta\hat{p}_2)^2\rangle_i]\right)$. Using Equation 19, it is apparent that

*Equation 22:* $\Delta^2 \equiv \sum_i \eta_i \left(a^2[\langle(\Delta\hat{x}_1)^2\rangle_i + \langle(\Delta\hat{p}_1)^2\rangle_i] + \frac{1}{a^2}[\langle(\Delta\hat{x}_2)^2\rangle_i + \langle(\Delta\hat{p}_2)^2\rangle_i]\right) \geq \sum_i \eta_i \left(a^2[1] + \frac{1}{a^2}[1]\right)$

or $\Delta^2 \geq \left(a^2 + \frac{1}{a^2}\right) \sum_i \eta_i$ and using $\sum_i \eta_i = 1$ in Equation 17

*Equation 23:* $\Delta^2 \geq \left(a^2 + \frac{1}{a^2}\right)$.

Summarizing so far, from Equation 20 and Equation 23, we have

*Equation 24:* $\langle(\Delta\hat{u})^2\rangle_\rho + \langle(\Delta\hat{v})^2\rangle_\rho \geq \left(a^2 + \frac{1}{a^2}\right) + \sum_i \eta_i \langle\hat{u}\rangle_i^2 - (\sum_i \eta_i \langle\hat{u}\rangle_i)^2 + \sum_i \eta_i \langle\hat{v}\rangle_i^2 - (\sum_i \eta_i \langle\hat{v}\rangle_i)^2 = \left(a^2 + \frac{1}{a^2}\right) + \sum_i \eta_i\{\langle\hat{u}\rangle_i^2 + \langle\hat{v}\rangle_i^2\} - (\sum_i \eta_i \langle\hat{u}\rangle_i)^2 - (\sum_i \eta_i \langle\hat{v}\rangle_i)^2$.

By the Cauchy-Schwarz inequality, we can write

$(\sum_i \eta_i)(\sum_i \eta_i \langle\hat{u}\rangle_i^2) \geq (\sum_i \eta_i |\langle\hat{u}\rangle_i|)^2$ or with $\sum_i \eta_i = 1$ in Equation 17, $(\sum_i \eta_i \langle\hat{u}\rangle_i^2) \geq (\sum_i \eta_i |\langle\hat{u}\rangle_i|)^2$ or

*Equation 25:* $-(\sum_i \eta_i |\langle\hat{u}\rangle_i|)^2 \geq -(\sum_i \eta_i \langle\hat{u}\rangle_i^2)$.

Note also that because of the absolute value sign, $(\sum_i \eta_i |\langle\hat{u}\rangle_i|)^2 \geq (\sum_i \eta_i \langle\hat{u}\rangle_i)^2$ and using Equation 25 for the second inequality in the following equation,

*Equation 26:* $-(\sum_i \eta_i \langle\hat{u}\rangle_i)^2 \geq -(\sum_i \eta_i |\langle\hat{u}\rangle_i|)^2 \geq -(\sum_i \eta_i \langle\hat{u}\rangle_i^2)$.

Similarly, we can show,

*Equation 27:* $-(\sum_i \eta_i \langle\hat{v}\rangle_i)^2 \geq -(\sum_i \eta_i |\langle\hat{v}\rangle_i|)^2 \geq -(\sum_i \eta_i \langle\hat{v}\rangle_i^2)$.

From Equation 26 and Equation 27, Equation 24 gives $\langle(\Delta \hat{u})^2\rangle_\rho + \langle(\Delta \hat{v})^2\rangle_\rho \geq \left(a^2 + \frac{1}{a^2}\right) + \sum_i \eta_i\{\langle \hat{u}\rangle_i^2 + \langle \hat{v}\rangle_i^2\} - (\sum_i \eta_i \langle \hat{u}\rangle_i)^2 - (\sum_i \eta_i \langle \hat{v}\rangle_i)^2 \geq \left(a^2 + \frac{1}{a^2}\right) + \sum_i \eta_i\{\langle \hat{u}\rangle_i^2 + \langle \hat{v}\rangle_i^2\} - (\sum_i \eta_i \langle \hat{u}\rangle_i^2) - (\sum_i \eta_i \langle \hat{v}\rangle_i^2)$ leaving only

*Equation 28:* $\langle(\Delta \hat{u})^2\rangle_\rho + \langle(\Delta \hat{v})^2\rangle_\rho \geq \left(a^2 + \frac{1}{a^2}\right)$

for separable states involving a pair of EPR-like operators $\hat{u}$ and $\hat{v}$. From Equation 28 we can conclude that if $\langle(\Delta \hat{u})^2\rangle_\rho + \langle(\Delta \hat{v})^2\rangle_\rho < \left(a^2 + \frac{1}{a^2}\right)$ then we have inseparable states. Hence the violation of Equation 28 gives us a sufficient condition for inseparability. In summary,

*Equation 29:* $\langle(\Delta \hat{u})^2\rangle_\rho + \langle(\Delta \hat{v})^2\rangle_\rho < \left(a^2 + \frac{1}{a^2}\right)$

for inseparable, entangled states.

### 2.3. Separability Condition for Entangled Particles, Tripartite Case

In [17], Loock and Furusawa worked out a condition similar to Duan's case for the tripartite case for partially and fully separable cases. We will show the relevant parts of the derivation of the fully separable case below.

We now look at the density operator $\rho$ of a completely separable tripartite state with modes 1, 2 and 3, namely $\rho = \sum_i \eta_i \rho_{i1} \otimes \rho_{i2} \otimes \rho_{i3}$ with $\eta_i \geq 0$ and $\sum_i \eta_i = 1$. Consider a generalization of Equation 18 for the tripartite case, $\hat{u} = h_1 \hat{x}_1 + h_2 \hat{x}_2 + h_3 \hat{x}_3$ and $\hat{v} = g_1 \hat{p}_1 + g_2 \hat{p}_2 + g_3 \hat{p}_3$ with real parameters $h_i$ and $g_i$. Using Equation 8, we expand $\langle(\Delta \hat{u})^2\rangle_\rho + \langle(\Delta \hat{v})^2\rangle_\rho$ and get [17]

*Equation 30:* $\langle(\Delta \hat{u})^2\rangle_\rho + \langle(\Delta \hat{v})^2\rangle_\rho = \sum_i \eta_i [T_{123i} + R_{123i}] + S_{ii}$

where we define

$T_{123i} \equiv h_1^2 \langle(\Delta \hat{x}_1)^2\rangle_i + h_2^2 \langle(\Delta \hat{x}_2)^2\rangle_i + h_3^2 \langle(\Delta \hat{x}_3)^2\rangle_i + g_1^2 \langle(\Delta \hat{p}_1)^2\rangle_i + g_2^2 \langle(\Delta \hat{p}_2)^2\rangle_i + g_3^2 \langle(\Delta \hat{p}_3)^2\rangle_i + 2h_1 h_2 (\langle \hat{x}_1 \hat{x}_2\rangle_i - \langle \hat{x}_1\rangle_i \langle \hat{x}_2\rangle_i) + 2g_1 g_2 (\langle \hat{p}_1 \hat{p}_2\rangle_i - \langle \hat{p}_1\rangle_i \langle \hat{p}_2\rangle_i)$,

$R_{123i} = 2h_1 h_3 (\langle \hat{x}_1 \hat{x}_3\rangle_i - \langle \hat{x}_1\rangle_i \langle \hat{x}_3\rangle_i) + 2h_2 h_3 (\langle \hat{x}_2 \hat{x}_3\rangle_i - \langle \hat{x}_2\rangle_i \langle \hat{x}_3\rangle_i) + 2g_1 g_3 (\langle \hat{p}_1 \hat{p}_3\rangle_i - \langle \hat{p}_1\rangle_i \langle \hat{p}_3\rangle_i) + 2g_2 g_3 (\langle \hat{p}_2 \hat{p}_3\rangle_i - \langle \hat{p}_2\rangle_i \langle \hat{p}_3\rangle_i)$ and

$S_{ii} = \sum_i \eta_i \langle \hat{u}\rangle_i^2 - (\sum_i \eta_i \langle \hat{u}\rangle_i)^2 + \sum_i \eta_i \langle \hat{v}\rangle_i^2 - (\sum_i \eta_i \langle \hat{v}\rangle_i)^2$.

As before in section 2.2, by the Cauchy-Schwarz inequality, we can show that $S_{ii} \geq 0$. Hence, Equation 30 becomes

*Equation 31:* $\langle(\Delta \hat{u})^2\rangle_\rho + \langle(\Delta \hat{v})^2\rangle_\rho = \sum_i \eta_i [T_{123i} + R_{123i}] + S_{ii} \geq \sum_i \eta_i [T_{123i} + R_{123i}]$.

Since modes 1, 2 and 3 are separable, $\langle \hat{x}_m \hat{x}_n\rangle_i = \langle \hat{x}_m\rangle_i \langle \hat{x}_n\rangle_i$ and $\langle \hat{p}_m \hat{p}_n\rangle_i = \langle \hat{p}_m\rangle_i \langle \hat{p}_n\rangle_i$. Clearly $R_{123i} = 0$ and Equation 31 becomes,

*Equation 32:* $\langle(\Delta \hat{u})^2\rangle_\rho + \langle(\Delta \hat{v})^2\rangle_\rho \geq \sum_i \eta_i T_{123i}$

with

*Equation 33:* $T_{123i} = h_1^2 \langle(\Delta \hat{x}_1)^2\rangle_i + h_2^2 \langle(\Delta \hat{x}_2)^2\rangle_i + h_3^2 \langle(\Delta \hat{x}_3)^2\rangle_i + g_1^2 \langle(\Delta \hat{p}_1)^2\rangle_i + g_2^2 \langle(\Delta \hat{p}_2)^2\rangle_i + g_3^2 \langle(\Delta \hat{p}_3)^2\rangle_i.$

Note that we can write $h_n^2 \langle(\Delta \hat{x}_n)^2\rangle_i + g_n^2 \langle(\Delta \hat{p}_n)^2\rangle_i = \langle(\Delta[h_n \hat{x}_n])^2\rangle_i + \langle(\Delta[g_n \hat{p}_n])^2\rangle_i$ and similar to Equation 19, and using $[\hat{x}_i, \hat{p}_j] = i\delta_{ij}$,

*Equation 34:* $h_n^2 \langle(\Delta \hat{x}_n)^2\rangle_i + g_n^2 \langle(\Delta \hat{p}_n)^2\rangle_i = \langle(\Delta[h_n \hat{x}_n])^2\rangle_i + \langle(\Delta[g_n \hat{p}_n])^2\rangle_i \geq |\langle[h_n \hat{x}_n, g_n \hat{p}_n]\rangle_i| = |h_n g_n \langle[\hat{x}_n, \hat{p}_n]\rangle_i| = |h_n g_n|.$

Hence from Equation 33 and Equation 34, $T_{123i} \geq |h_1 g_1| + |h_2 g_2| + |h_3 g_3|$. Equation 32 becomes

*Equation 35:* $\langle(\Delta \hat{u})^2\rangle_\rho + \langle(\Delta \hat{v})^2\rangle_\rho \geq \sum_i \eta_i \{|h_1 g_1| + |h_2 g_2| + |h_3 g_3|\}$ and with $\sum_i \eta_i = 1$,

$\langle(\Delta \hat{u})^2\rangle_\rho + \langle(\Delta \hat{v})^2\rangle_\rho \geq |h_1 g_1| + |h_2 g_2| + |h_3 g_3|.$

Hence similar to the bipartite case,

*Equation 36:* $\langle(\Delta \hat{u})^2\rangle_\rho + \langle(\Delta \hat{v})^2\rangle_\rho < \sum_i \eta_i \{|h_1 g_1| + |h_2 g_2| + |h_3 g_3|\} = |h_1 g_1| + |h_2 g_2| + |h_3 g_3|$

for inseparable, entangled states.

## 3. The Generalized Uncertainty Principle and Entanglement

We now calculate the GUP corrected equations. From Equation 1, we see that the GUP-corrected commutation relation of Equation 5 are given by

*Equation 37:* $[Q_i, P_j] = i\hbar \delta_{ij}(1 + \beta P_i^2), [Q_i, Q_j] = [P_i, P_j] = 0.$

### 3.1. Modified Uncertainty Relations for Identical, Entangled Particles with GUP correction

It is apparent from the calculations in subsection 2.1 above that, what will change in the relations above is the right hand side of the inequality due to the extra term proportional to the $\beta$ parameter. Using *Equation 37* in Equation 6 for $N = 2$ particles, results in $[Q, P] = 2i\hbar + i\hbar\beta[P_1^2 + P_2^2]$. Hence we get (to order $\beta$ only)

*Equation 38:* $(\Delta Q)^2 (\Delta P)^2 \geq \hbar^2 + \hbar^2 \beta [\langle P_1^2\rangle + \langle P_2^2\rangle].$

The above equation replaces Equation 7 for $N = 2$ particles. Let us rewrite Equation 38 in terms of $\Delta P_i^2$ with $i = 1,2$. From Equation 8, we can write $\langle P_i^2 \rangle = (\Delta P_i)^2 + \langle P_i \rangle^2$. Hence Equation 38 becomes $(\Delta Q)^2 (\Delta P)^2 \geq \hbar^2 + \hbar^2 \beta [(\Delta P_1)^2 + (\Delta P_2)^2 + \langle P_1 \rangle^2 + \langle P_2 \rangle^2] \geq \hbar^2 + \hbar^2 \beta [(\Delta P_1)^2 + (\Delta P_2)^2]$. It follows that

*Equation 39:* $(\Delta Q)^2 (\Delta P)^2 \geq \hbar^2 + \hbar^2 \beta [(\Delta P_1)^2 + (\Delta P_2)^2].$

One can retrace the steps above using Equation 39 to get the GUP-corrected Equation 15.

*Equation 40:* $[(\Delta Q_1)^2 + (\Delta Q_2)^2] \times [(\Delta P_1)^2 + (\Delta P_2)^2] \geq \frac{\hbar^2}{4} + \frac{\hbar^2}{4}\beta((\Delta P_1)^2 + (\Delta P_2)^2)$

which apparently reduces to Equation 15 when $\beta = 0$. With the special case when $\Delta Q_1 = \Delta Q_2$ and $\Delta P_1 = \Delta P_2$, we get $(\Delta Q_i)^2(\Delta P_i)^2 \geq \frac{\hbar}{16}(1 + 2\beta(\Delta P_i)^2)$ which can be rewritten as (to the order $\beta$)

*Equation 41:* $\Delta Q_i \Delta P_i \geq \frac{\hbar}{4} + \Delta_{GUP}$, with $\Delta_{GUP} \equiv \frac{\hbar}{4}\beta(\Delta P_i)^2 \geq 0$, $i = 1,2$.

Rigolin's motivation [14, 15] in deriving Equation 16 is to give a possible alternative explanation for the experimental results of Kim and Shih [28] which seem to imply a violation of the uncertainty principle in that $\Delta y \Delta p_y < \hbar$. The presence of the GUP correction puts a constraint upon which certain states of two identical and entangled particles may or may not violate the uncertainty principle. Using the form of the HUP in section 1, $\Delta x \Delta p \geq \hbar/2$, if $\Delta_{GUP} \geq \frac{\hbar}{4}$ or $\beta(\Delta P_i)^2 \geq 1$ then there is no disagreement with the uncertainty principle while $\Delta_{GUP} \leq \frac{\hbar}{4}$ or $\beta(\Delta P_i)^2 \leq 1$ will cause a disagreement from the uncertainty principle.

From [28] and [29], one can estimate the $\Delta P_i$ of the entangled photon along the $y$ axis to be $\Delta P_i \sim \frac{\hbar}{2\Delta y}$ where $\Delta y = 0.16$ mm. Apparently, with a small $\beta$, $\beta(\Delta P_i)^2 \leq 1$ and hence this causes a disagreement with the uncertainty principle as similar to Rigolin's result. However, since the $\Delta_{GUP}$ is positive in Equation 41, GUP effects increase the lower bound of the product of the variances and <u>lessens</u> the disagreement with the HUP for identical entangled particles. In [15] it was shown that as the number $N$ of identical entangled particles increase, the lower bound of the product of the position and momentum uncertainties decreases approaching zero as $N$ becomes very large, moving towards classicality. However, the presence of the GUP correction reduces the tendency towards classicality (hence making it more quantum mechanical) as compared to the non-GUP case for a large number of entangled identical particles.

### 3.2. Separability Condition for Entangled Particles, Bipartite Case with GUP Correction

From Equation 1, we have $[x_i, p_i] = i\hbar(1 + \beta p_i^2)$. Using this equation, the GUP-corrected Equation 19 is (with $\hbar = 1$), $\langle(\Delta \hat{x}_j)^2\rangle_i + \langle(\Delta \hat{p}_j)^2\rangle_i \geq |\langle[\hat{x}_j, \hat{p}_j]\rangle_i| = |1 + \beta\langle\hat{p}_j^2\rangle_i|$. Since the quantity inside the absolute value is positive, we have,

*Equation 42:* $\langle(\Delta \hat{x}_j)^2\rangle_i + \langle(\Delta \hat{p}_j)^2\rangle_i \geq 1 + \beta\langle\hat{p}_j^2\rangle_i$.

From Equation 42, we get the GUP-corrected Equation 22 as $\Delta^2 \equiv \sum_i \eta_i \left(a^2[\langle(\Delta \hat{x}_1)^2\rangle_i + \langle(\Delta \hat{p}_1)^2\rangle_i] + \frac{1}{a^2}[\langle(\Delta \hat{x}_2)^2\rangle_i + \langle(\Delta \hat{p}_2)^2\rangle_i]\right) \geq \sum_i \eta_i \left(a^2[1 + \beta\langle\hat{p}_1^2\rangle_i] + \frac{1}{a^2}[1 + \beta\langle\hat{p}_2^2\rangle_i]\right)$ or

*Equation 43:* $\Delta^2 \geq \left(a^2 + \frac{1}{a^2}\right) + \beta f(\hat{p}_1, \hat{p}_2)$ where $f(\hat{p}_1, \hat{p}_2) \equiv \sum_i \eta_i \left[a^2\langle\hat{p}_1^2\rangle_i + \frac{1}{a^2}\langle\hat{p}_2^2\rangle_i\right]$

Hence the GUP-corrected Equation 28 is

*Equation 44:* $\langle(\Delta\hat{u})^2\rangle_\rho + \langle(\Delta\hat{v})^2\rangle_\rho \geq \left(a^2 + \frac{1}{a^2}\right) + \Delta_{GUP}$ where $\Delta_{GUP} \equiv \sum_i \eta_i \left[a^2\langle\hat{p}_1^2\rangle_i + \frac{1}{a^2}\langle\hat{p}_2^2\rangle_i\right] \geq 0$.

The GUP-corrected Equation 29 is

*Equation 45:* $\langle(\Delta\hat{u})^2\rangle_\rho + \langle(\Delta\hat{v})^2\rangle_\rho < \left(a^2 + \frac{1}{a^2}\right) + \Delta_{GUP}$

for inseparable states where $\Delta_{GUP}$ is given in *Equation 44*. Comparing *Equation 29* and *Equation 45*, with $\Delta_{GUP} \geq 0$, the upper bound for inseparability is increased, increasing the range for inseparability and hence enhancing the entanglement of two-party continuous variable states.

### 3.3. Separability Condition for Entangled Particles, Tripartite Case with GUP Correction

Turning now to the tripartite case of section 2.3, from Equation 34,

*Equation 46:* $h_n^2\langle(\Delta\hat{x}_n)^2\rangle_i + g_n^2\langle(\Delta\hat{p}_n)^2\rangle_i = \langle(\Delta[h_n\hat{x}_n])^2\rangle_i + \langle(\Delta[g_n\hat{p}_n])^2\rangle_i \geq |\langle[h_n\hat{x}_n, g_n\hat{p}_n]\rangle_i| = |h_n g_n\langle[\hat{x}_n, \hat{p}_n]\rangle_i| = |h_n g_n\langle 1 + \beta\hat{p}_n^2\rangle_i| = |h_n g_n(1 + \beta\langle\hat{p}_n^2\rangle_i)|$.

Hence from *Equation 33* and *Equation 46*, we have $T_{123i} \geq |h_1 g_1(1 + \beta\langle p_1^2\rangle_i)| + |h_2 g_2(1 + \beta\langle p_2^2\rangle_i)| + |h_3 g_3(1 + \beta\langle p_3^2\rangle_i)|$. From *Equation 32*, we have

*Equation 47:* $\langle(\Delta\hat{u})^2\rangle_\rho + \langle(\Delta\hat{v})^2\rangle_\rho \geq \sum_i \eta_i\{|h_1 g_1(1 + \beta\langle p_1^2\rangle_i)| + |h_2 g_2(1 + \beta\langle p_2^2\rangle_i)| + |h_3 g_3(1 + \beta\langle p_3^2\rangle_i)|\}$

for completely separable states. Therefore, for inseparable states,

*Equation 48:* $\langle(\Delta\hat{u})^2\rangle_\rho + \langle(\Delta\hat{v})^2\rangle_\rho < \sum_i \eta_i\{|h_1 g_1(1 + \beta\langle p_1^2\rangle_i)| + |h_2 g_2(1 + \beta\langle p_2^2\rangle_i)| + |h_3 g_3(1 + \beta\langle p_3^2\rangle_i)|\}$.

Comparing Equation 48 and Equation 36, the upper bound for inseparability is increased, increasing the range for inseparability and hence enhancing the entanglement of three-party continuous variable states.

## 4. Conclusions

By a straightforward application of the modified commutation relation (Equation 1), we have shown how the generalized uncertainty principle or the presence of minimal length affect entanglement. In all three cases, the bounds were increased by GUP effects in such a way as to make the quantum effects more pronounced. In the first case an increase in the lower bound makes the disagreement with the uncertainty principle less such that there is less classicality when the number of identical entangled particles becomes large. In the second and third cases, an increase in the upper bound of the inseparability condition increases the range of inseparability and hence enhances entanglement. These results are in agreement with reference [30] in which it was shown that nonclassicality and entanglement are enhanced by increasing the noncommutativity of the underlying space (an alternate way of introducing minimal length similar to the GUP). Similarly, in ref [31], it is shown that the entanglement of Gaussian states can be induced by a noncommutative phase space scenario.

In this paper, we confined ourselves in the study of GUP effects on continuous variable systems with the variance form of the uncertainty principle ($\Delta x \Delta p \geq \hbar/2$). Further studies on the effect of minimal length and GUP on continuous variable systems can be done for multipartite systems [17, 20, 22].

An interesting study of the effects of minimal length and GUP on entanglement can be made via the entropic uncertainty principle [23-25, 33-35]. This will be the topic of a future publication.

## Acknowledgements

The authors would like to thank Rafael Sepulveda of Houston Baptist University for his work on the initial stages of this paper.